\documentclass[runningheads]{llncs}
\usepackage[noadjust]{cite}
\usepackage{amsmath,amssymb,amsfonts}
\usepackage{algorithm}
\usepackage{algpseudocode}
\usepackage{graphicx}
\usepackage{textcomp}
\usepackage{subcaption}
\usepackage{xcolor}
\usepackage[T1]{fontenc}
\usepackage{multicol}
\usepackage{multirow}
\usepackage{placeins}
\usepackage{setspace}
\usepackage{enumitem}
\usepackage{makecell}
\usepackage{caption}
\usepackage{placeins} 
\usepackage{authblk}
\usepackage{orcidlink}

\titlerunning{FLaaS for Hierarchical Edge Networks with Heterogeneous Models}

\setlist[itemize]{topsep=0pt, partopsep=0pt, leftmargin=0pt}


\newcommand{\numberedauthor}[2]{#1\textsuperscript{#2}}
\begin{document}
\title{Federated Learning as a Service for Hierarchical Edge Networks with Heterogeneous Models}

\author{  \numberedauthor{Wentao Gao}{1}\orcidlink{0000}, 
  \numberedauthor{Omid Tavallaie}{1,2}\orcidlink{0000-0002-3367-1236}, 
  \numberedauthor{Shuaijun Chen}{1}\orcidlink{0009-0001-4944-3406},
  \numberedauthor{Albert Zomaya}{1}\orcidlink{0000-0002-3090-1059}
  }
  
\institute{\numberedauthor{The University of Sydney}{1},
        \numberedauthor{University of Oxford}{2}}

\institute{School of Computer Science, The University of Sydney, Australia \and Department of Engineering Science, University of Oxford, United Kingdom
\email\{wentao.gao, shuaijun.chen, albert.zomaya\}@sydney.edu.au, omid.tavallaie@eng.ox.ac.uk}

\maketitle    

\begin{abstract}
Federated learning (FL) is a distributed Machine Learning (ML) framework that is capable of training a new global model by aggregating clients' locally trained models without sharing users' original data. Federated learning as a service (FLaaS) offers a privacy-preserving approach for training machine learning models on devices with various computational resources. Most proposed FL-based methods train the same model in all client devices regardless of their computational resources. However, in practical Internet of Things (IoT) scenarios, IoT devices with limited computational resources may not be capable of training models that client devices with greater hardware performance hosted. Most of the existing FL frameworks that aim to solve the problem of aggregating heterogeneous models are designed for Independent and Identical Distributed (IID) data, which may make it hard to reach the target algorithm performance when encountering non-IID scenarios. To address these problems in hierarchical networks, in this paper, we propose a heterogeneous aggregation framework for hierarchical edge systems called HAF-Edge. In our proposed framework, we introduce a communication-efficient model aggregation method designed for FL systems with two-level model aggregations running at the edge and cloud levels. This approach enhances the convergence rate of the global model by leveraging selective knowledge transfer during the aggregation of heterogeneous models. To the best of our knowledge, this work is pioneering in addressing the problem of aggregating heterogeneous models within hierarchical FL systems spanning IoT, edge, and cloud environments. We conducted extensive experiments to validate the performance of our proposed method. The evaluation results demonstrate that HAF-Edge significantly outperforms state-of-the-art methods.

\keywords{Federated Learning as a Service (FLaaS), Model Heterogeneity, Edge Computing, IoT Networks}
\end{abstract}
\section{Introduction}
\begin{figure}[t]
  \centering
    \includegraphics[width=1\linewidth, height= 42 mm]{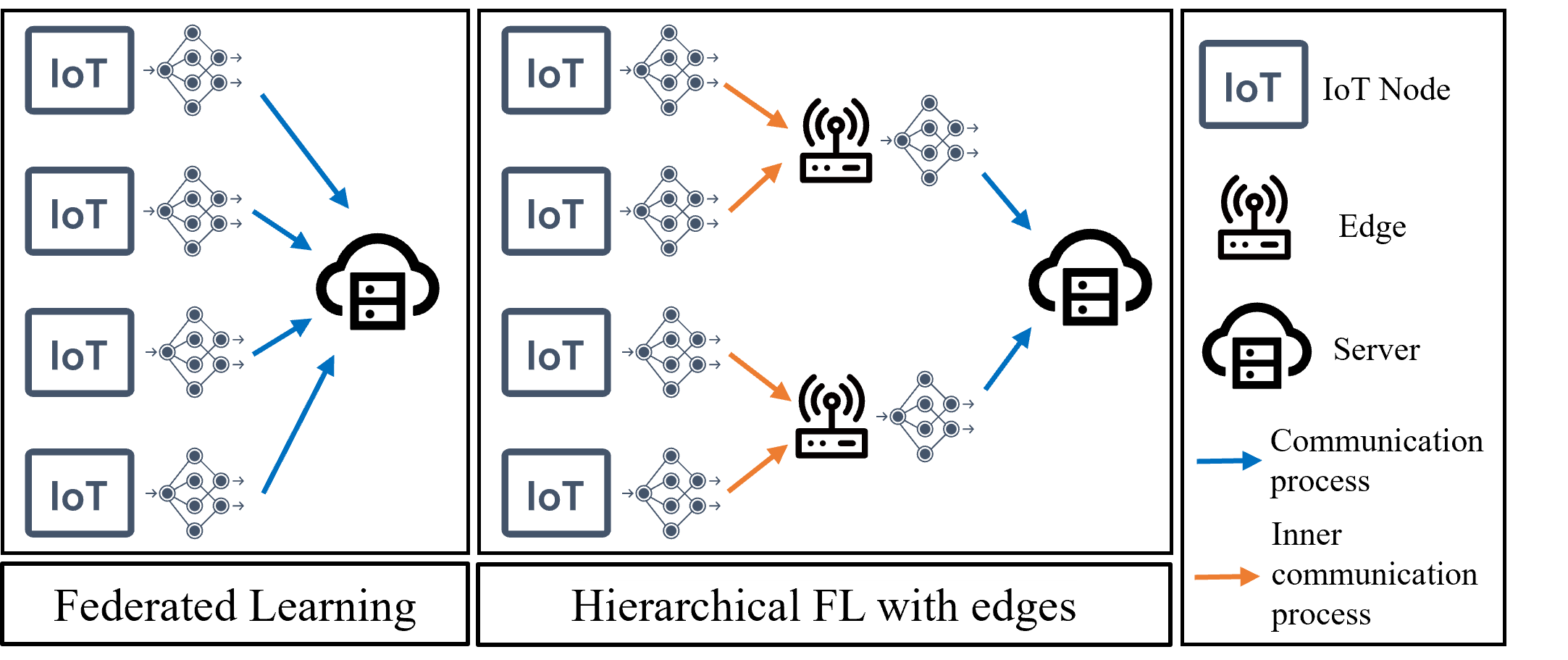}
  \caption{Comparison between vanilla FL and Hierarchical FL}
  \label{fig:fl_hfl_comparison}
\end{figure}

Federated learning (FL) \cite{fl} is a distributed learning algorithm designed for privacy-aware applications. Without compromising the user's privacy, FL aggregates trained local models to build a new global model \cite{nazemi_privacy}. In each training round, all client devices receive the global model from the cloud server, train it using local data, and send the updated model back to the server for building a new global model in the aggregation process (Figure \ref{fig:fl_hfl_comparison}). Compared with traditional centralized training, FL alleviates concerns about direct data violation \cite{privcayconcern}, as the client's original data is never transmitted to the server in the training process. In recent years, Machine Learning as a Service (MLaaS) has gained widespread attention, proved by the increasing demand for cloud-based machine learning platforms and large-scale analytics across various industries \cite{276938}. However, the concerns of widely collecting privacy data have also grown \cite{hesamifard2018privacy}. FL, as a decentralized machine learning, has the capability to face the challenges of privacy concerns. In 2020, Nicolas et al. proposed Federated Learning as a Service (FLaaS) for permission and privacy management \cite{flaas}. In recent years, FL has been applied in various applications such as healthcare systems \cite{healthcare}, prediction maintenance for Internet of Things (IoT) devices \cite{prediction}, and edge networks \cite{tsc1,tsc2}. 

Data heterogeneity among client devices is one of the major challenges in FL \cite{challenge}. In practical FL scenarios, the client's data is highly dependent on user behavior. As a result, training data on clients' devices may have significantly different distributions \cite{hsieh2020non}. This phenomenon is called non-Independent and Identical (non-IID) distributed data, which has a significant impact on the global model's performance \cite{li2019convergence,chen2024optimizationfederatedlearningsclient}. Device heterogeneity in the computational hierarchy is another existing problem of FL. Most proposed FL frameworks require all participants to use the same model architecture regardless of their computational resources. However, in practical FL scenarios such as Google Gboard \cite{xu-etal-2023-federated}, mobile clients could have various hardware configurations \cite{273723}. In this setting, running the same model on all client devices is challenging, if it is not impossible. Devices with low computational resources are incapable of finishing the training process on time \cite{273723}, which increases the average time for a training round as the server can perform the model aggregation when it receives trained models from all devices. 

To address these research problems, various methods have been proposed to aggregate heterogeneous models within conventional Federated Learning (FL) systems \cite{meng2022improving}. However, most of these methods cannot reach the target performance when encountering non-IID data \cite{li2020federated}. In this paper, we introduce an aggregation method for heterogeneous models in three-level hierarchical FL frameworks (IoT/edge/cloud) called \textbf{HAF-Edge}. In this computational hierarchy, IoT devices are clustered based on some specific requirements (data/hardware configurations), such as the facilities in a smart city or industry of IoT. Each cluster is connected to an edge server that receives the trained models and aggregates them to create an edge-aggregated model. All edge servers are connected to a cloud server that receives edge-aggregated models with different structures. The cloud server performs the second level of aggregation by using knowledge sharing and based on similarity in structures of different models to build a global model for each edge server. HAF-Edge not only leverages the advantages of hierarchical federated learning, potentially reducing the need for the Internet to aggregate trained IoT models at the edge \cite{hierarchicalcommunication}, but also addresses the challenge of aggregating edge models with different settings at the cloud level. By connecting a cluster of IoT nodes with the same setting to an edge server using local communication technologies (e.g., Bluetooth or Zigbee), the need for using the Internet in creating edge-aggregated models is eliminated \cite{hierarchicalcommunication} (reduces communication cost). HAF-Edge addresses the challenge of \textbf{aggregating heterogeneous models at the cloud by adopting a strategy called MaxCommon} \cite{flexifed} in hierarchical edge networks. At the cloud level, we extract parts of the knowledge from different model settings to enhance the global model's aggregation process. Our experimental results demonstrate that \textbf{models trained with IID data tend to have a greater distance from the global model compared to scenarios where non-IID data is used}. Based on this observation, we propose a distance-based weighting aggregation approach and apply it to the aggregation process at the edge level to improve performance for applications with non-IID data. The main contributions of our work are summarized as follows:
\begin{itemize}
\item 1) To the best of our knowledge, HAF-Edge is the first attempt to resolve the challenge of aggregating heterogeneous models in hierarchical federated learning architecture with two levels of aggregations (edge and the cloud).
\item 2) A distance-based weighting aggregation approach is proposed, which makes HAF-Edge enables to achieve a better performance facing non-IID data compared to the state-of-the-art FL frameworks.
\item 3) We perform extensive sets of experiments on two public datasets (MNIST \cite{MNIST} and FMNIST \cite{FMNIST}). The evaluation results show that HAF-Edge achieved a better performance compared to state-of-the-art methods.
\end{itemize}

\section{Related Works}
FL \cite{fl} is a decentralized model training method that was proposed for training neural networks in a distributed fashion without sharing the user's data. However, the existence of non-IID data among client devices in practical FL scenarios is one of the most important problems that reduces the performance of the global model \cite{non-iid}. In vanilla FedAvg, only data volume is considered in the aggregation process. As a result, in practical FL scenarios with non-IID data, regardless of data distribution, the same weighting coefficient is considered for trained models of two client devices that had data with the same size but with completely different labels. Broadly aggregating the models trained with low-quality data can have a serious impact on the performance of the global model in FL \cite{fair}. Some proposed methods start to shift the focus to the quality of client data or models. FAIR proposed in \cite{fair} improves the performance of the global model by filtering the local models with low quality.

In practical IoT FL scenarios, the expected local model architecture may differ among client devices due to heterogeneous hardware configurations, tasks, and personal demands \cite{challenge1,challenge,challenge2}. To address these challenges, FedMD \cite{FedMD} has been proposed based on knowledge distillation \cite{KnowledgeDistillation}. However, in the methods using knowledge distillation, a public dataset is created, which can compromise client data privacy. In contrast to knowledge distillation, some methods aggregate common layers of heterogeneous models to share learned features among participants, such as MaxCommon \cite{flexifed} and Rank-Based Lora Aggregation \cite{chen2024rblarankbasedloraaggregationfinetuningheterogeneous}. MaxCommon targets heterogeneous model architectures and refines the aggregation of local updates to specific layers rather than entire models. By aggregating as many common layers as possible from different clients' models, MaxCommon maximizes knowledge sharing. The method employs data quantity-based aggregation to merge common layers from various clients' updates, ensuring that a client's model with a larger training dataset contributes more to the aggregated common layers. However, this approach does not consider the distribution of local training data.

Hierarchical federated learning (Figure \ref{fig:fl_hfl_comparison}) is realized by introducing hierarchical clustering steps into FL \cite{hierarchicalfl1}, where client devices are clustered according to specific requirements to avoid sending trained models directly to the cloud server for aggregation. A common three-level structure in FL is the client-edge-cloud model \cite{hierarchicalfl}. In this model, edge servers aggregate the received local updates from clients and then send the aggregated updates to the cloud server, which subsequently aggregates the updates from the edges to create a new global model. In hierarchical FL, IoT devices are connected to edge servers rather than directly to the cloud server (the cloud server communicates only with edge servers). One apparent advantage of hierarchical FL compared to conventional FL networks is the reduction in communication costs. By aggregating local updates within clusters first, the number of updates sent to the cloud server is minimized. Communication costs between IoT devices and edge servers are typically lower than that for the communication between IoT devices and the cloud server as local network communication technologies (e.g., Bluetooth or Zigbee) are utilized to connect IoT devices to edge servers \cite{hierarchicalcommunication}. Additionally, hierarchical federated learning can reduce communication latency as the aggregation process for the local network is placed at the edge \cite{abad2020hierarchical}.

\section{Framework Design}

\begin{figure}[t]
  \centering
    \includegraphics[width=1\linewidth, height= 42 mm]{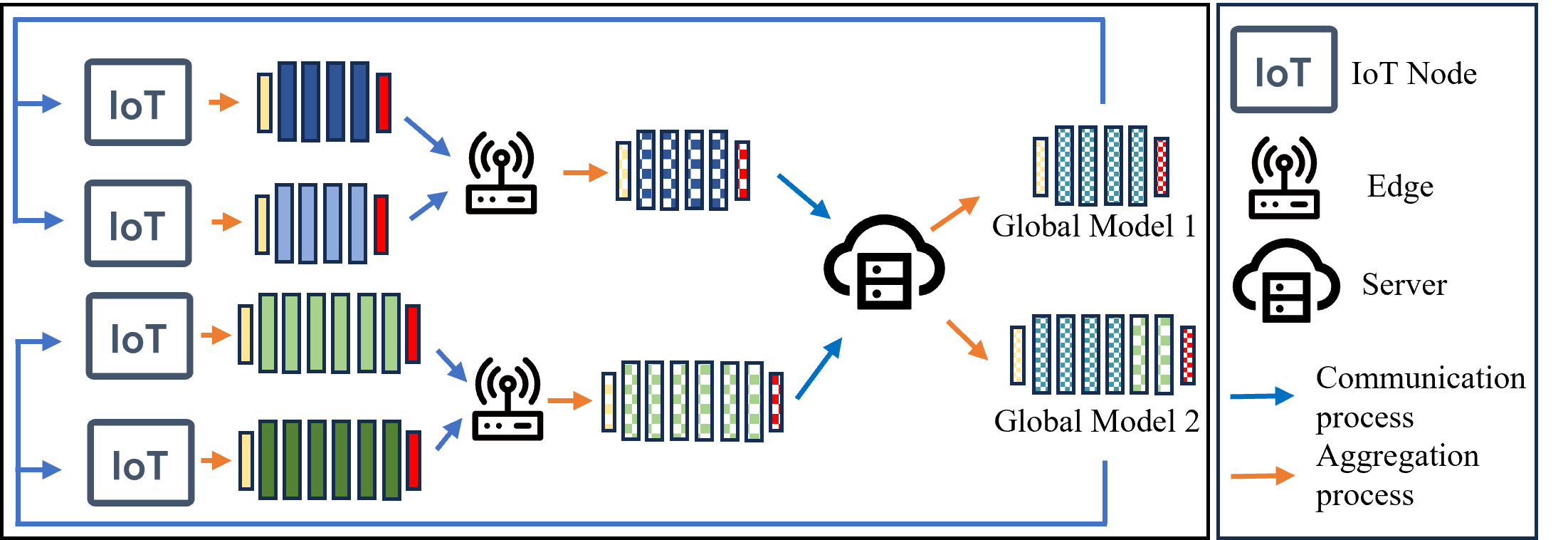}
  \caption{Entire process of HAF-Edge, where orange and blue arrow represents inner aggregation and communication process, respectively.}
  \label{fig:entire_process}
\end{figure}

By employing HAF-Edge, client devices are clustered based on their model architectures to ensure that IoT devices with identical model architectures are grouped into the same cluster. Each cluster of IoT devices is connected to an edge server which aggregates local models and forwards the edge-aggregated model to the cloud server. Then, edge-aggregated models with heterogeneous architectures are aggregated in the cloud by using the MaxCommon strategy. Subsequently, the cloud server distributes the updated global models with distinct architectures to corresponding edge servers. Edge servers then propagate these new global models to the IoT devices within their clusters, guaranteeing synchronized updates across edge servers and client devices (Figure \ref{fig:entire_process}).   

The distance-based weighting is implemented at the edge server to enable the weights of the contribution from client devices depending on the quality of training data instead of quantity so that \textbf{HAF-Edge can achieve better performance in the scenarios involving non-IID data compared to FedAvg and MaxCommon Strategy}. The distance-based weighting guarantees that local models trained with IID data have a greater contribution to the edge-aggregated model. The detailed steps of HAF-Edge for one completed round after clustering client devices based on their model architectures are:
\begin{itemize}

\item First step, in each cluster, the client devices train the models with their local data and send the trained local models to the edge server.

\item Second step, each edge server aggregates the trained local models from the client devices and then forwards the edge-aggregated model to the cloud server.

\item Third step, the cloud server applies MaxCommon strategy on edge-aggregated models with different architectures to \textbf{create several new heterogeneous global models}.

\item Fourth step, the cloud server transmits new global models to the matching edge servers. Then each edge server distributes the received new global model to the connected client devices.
\end{itemize}

\subsection{Distance-based Weighting}

\begin{figure}[t]
\begin{subfigure}{0.5\textwidth}
\includegraphics[width=0.9\linewidth, height=36 mm]{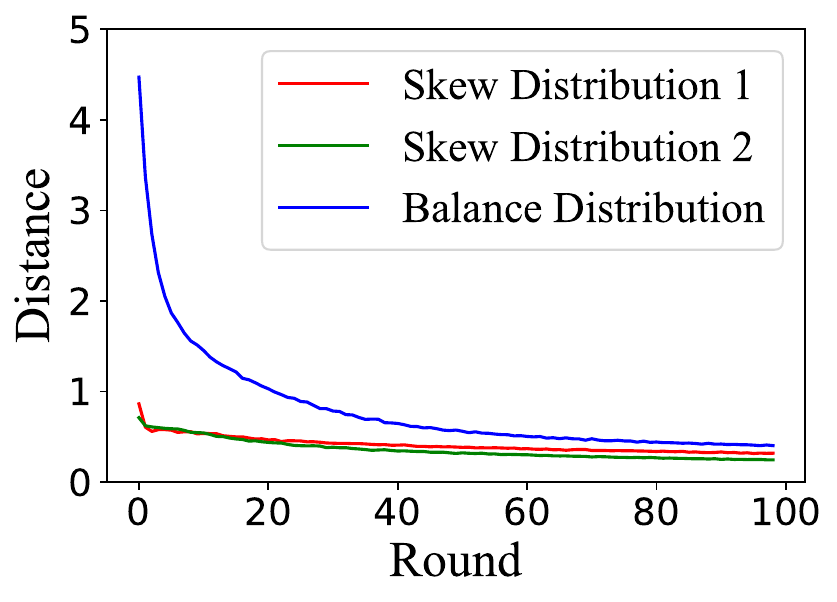} 
\caption{Configuration 1}
\label{fig:2non_IID and 1IID}
\end{subfigure}
\begin{subfigure}{0.5\textwidth}
\includegraphics[width=0.9\linewidth, height=36 mm]{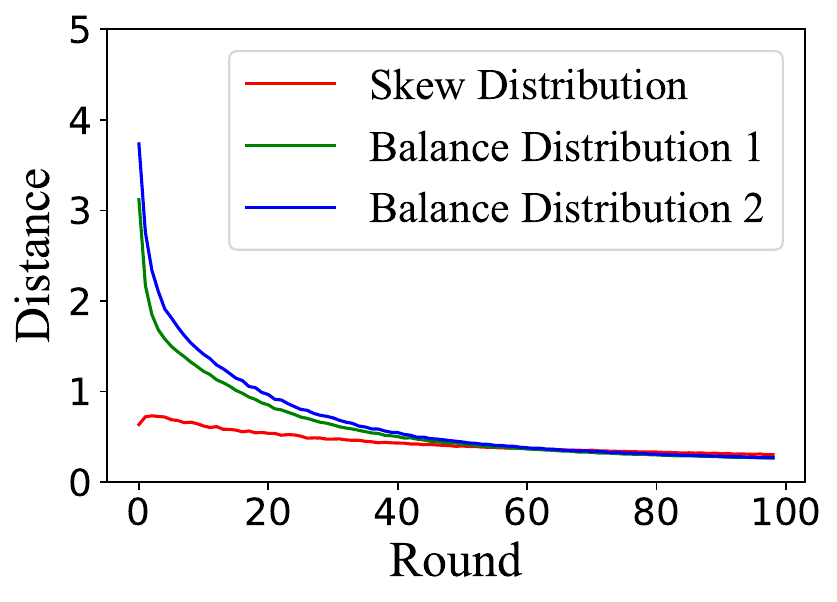}
\caption{Configuration 2}
\label{fig:1non_IID and 2IID}
\end{subfigure}
\caption{The changing curves of Euclidean distance between client model and global model per communication round on three clients with varied data distribution.}
\label{fig:distance}
\end{figure}


In HAF-Edge, we use the Euclidean distance-based weighting method to reduce the impact of non-IID data. In FL, under the non-IID scenario, the Euclidean distance of clients trained with skewer data distribution is lower than the models trained with balanced distributed data due to the neural network having to update more weights to capture more detailed client data features. We conducted an experiment with two different configurations:

\begin{itemize}
\item Configuration 1: In total three participants, one participant has balanced distributed data while the rest have skewed distributed data, data volume are the same across these devices.

\item Configuration 2: In total three participants, two participants have balanced distributed data while the other has skewed distributed data, data volume are the same across these devices.
\end{itemize}
Configuration 1 represents the scenario in which the number of client devices with non-IID data exceeds the number of client devices with IID data in the aggregation process. Conversely, configuration 2 indicates that there are more client devices with IID data. The Euclidean distance metric can be employed to quantify the disparity between the two models. In the Euclidean distance \cite{normdistance}:
\begin{equation}
    d(p,q) = \sqrt{(p_1-q_1)^2 + (p_2-q_2)^2 + (p_3-q_3)^2 + \cdots + (p_n-q_n)^2},
\end{equation}
where \(d(p,q)\) is the Euclidean distance between \(p\) and \(q\), \(p\) and \(q\) are matrices in n dimensions. FedAvg is utilized to aggregate three local models from client devices in both Configuration 1 and Configuration 2. The distance between the local models of three client devices and the aggregated global model from the previous round is calculated using the Euclidean distance metric.

Figure \ref{fig:distance} shows the changing curve of the Euclidean distance between client models and the global model by communication rounds. In both configurations, the Euclidean distance to the global model of local models trained with skewed data is lower than the local models from the client devices having balanced data. The distance between the local models trained with IID data to the global model decreases from \textbf{3.1-4.4 to 0.27-0.4}, and for the models with non-IID data decreases from \textbf{0.5-0.9 to 0.24-0.31}, which \textbf{verified our insight that regardless of whether the client's data is evenly distributed, the model trained with data in less non-IIDness has a higher Euclidean distance} to the global model compared with the models trained with higher data non-IIDness. Based on this observation, we propose a Euclidean distance-based weighting aggregation:

\begin{equation}
    \frac{d(L_j^t,G_i^{t-1})}{\sum_{k \in K_i}d(L_k^t,G_i^{t-1})},
    \label{eq:weight_assign}
\end{equation}
where \(K_i\) is the set of client devices connected with edge server \(i\), client \(j\) belongs to \(K_i\). \(L_j^t\) is the local model from client \(j\) at round \(t (t>1)\). \(G_i^{t-1}\) is the global model for edge server \(i\) (the aggregated model from edge server \(i\) contains \(i\) layers in total) at round \(t-1 (t>1)\). If local models are trained with data in less non-IIDness, by applying Eq.\ref{eq:weight_assign} can assign these models a higher weight coefficient for aggregation.

\subsection{Edge Aggregation}

\begin{table}[t]
    \caption{Declaration of notations}
    \centering
    \begin{tabular}{|c|c|}
    \hline
    Notation & Definition \\
    \hline
    \(L_k^t\) & A local model from client \(k\) at round \(t\) \\
    \hline
    \(E_i^t\) & \makecell{An edge-aggregated model from edge server \(i\) at round \(t\) \\ (the aggregated model from edge server \(i\) contains \(i\) layers in total)}\\
    \hline
    \(N_k\) & The total number of data used for training in client \(k\) \\
    \hline
    \(l_i^j\) & Layer \(j\) from \(E_i\) \\
    \hline
    \(X_i\) & A set of edge servers whose aggregated models contain layer \(i\) \\
    \hline
    \(K_i\) & The set of client devices connected to edge server \(i\) \\
    \hline
    \(Gl_i^t\) & A global layer \(i\) at round \(t\) \\
    \hline
    \(G_i^t\) & A global model for edge server \(i\) at round \(t\) \\
    \hline
    \end{tabular}
    \label{tab:my_label}
\end{table}

The IoT devices connected to the same edge server have an identical model architecture. The model-level aggregation can be used at edge servers. In the initial round, the cloud server did not generate the aggregated global models. Hence, the aggregation based on the quantity of data is implemented at edge servers for the first round. The edge-aggregated model \(E_i^1\) at the first communication round from edge server \(i\) is shown as:

\begin{equation}    
    E_i^1 = \sum_{k \in K_i} \frac{N_k}{\sum_{k \in K_i} N_k} L_k^1, 
\end{equation}
where \(N_k\) is the total number of data used to train the local model in the client \(k\). After receiving edge-aggregated models, the cloud server creates new global models with different architectures by aggregating common layers from various edge-aggregated models. New global models are distributed to matching edge servers. The edge server propagates the received new global model to connected client devices. Hence, starting from the second round, the distance-based weighting aggregation is applied at the edge server. The edge-aggregated model \(E_i^t\) at round \(t (t>1)\) from edge server \(i\) is:
\begin{equation}    
    E_i^t = \sum_{k \in K_i} \frac{d(L_k^t,G_i^{t-1})}{\sum_{k \in K_i}d(L_k^t,G_i^{t-1})} L_k^t.
\end{equation}
Algorithm \ref{alg:alg1} shows the aggregation process at edge servers.

\begin{algorithm}[t]
\setstretch{1.5}
    \caption{ \textbf{The Aggregation at Edge Servers.} \(K_i\) is the set of client devices connected with edge server \(i\). \(L_k^t\) is the local model from client \(k\) at round \(t\). \(N_k\) is the number of data used for training in client \(k\). \(G_i^t\) is the global model from the cloud server for edge server \(i\) at round \(t\). \(E_i^t\) is the edge-aggregated model from edge server \(i\) at round \(t\). \(t\) is the round counter.} \label{alg:alg1}
    \begin{algorithmic}[1]
            \If{$t=1$}
                \State $E_i^t = \sum_{k \in K_i} \dfrac{N_k}{\sum_{k \in K_i} N_k} L_k^t$
            \Else
                \State $E_i^t = \sum_{k \in K_i} \dfrac{d(L_k^t,G_i^{t-1})}{\sum_{k \in K_i}d(L_k^t,G_i^{t-1})} L_k^t$
            \EndIf
        \State \Return $E_i^t$
    \end{algorithmic}
\end{algorithm} 
\subsection{Cloud Aggregation}
MaxCommon strategy is adopted to aggregate the models from different edge servers at the cloud server. MaxCommon strategy is a layer-level aggregation. Hence we define a function:

\begin{equation}
    l_j^i = Extract(E_i^t,j),
\end{equation}
to extract layer \(j\) from edge-aggregated model \(E_i^t\). Then, the cloud server aggregates the common layers extracted from distinct edge-aggregated models:

\begin{equation}
    Gl_i^t = \sum_{x\in X_i} \frac{\sum_{k \in K_x} N_k}{\sum_{x\in X_i} \sum_{k \in K_x} N_k}l_i^x,
\end{equation}
where \(Gl_i^t\) is the global layer \(i\) after aggregating at round \(t\), \(X_i\) is a set of edge servers whose aggregated models contain layer \(i\). \(K_x\) is the set of client devices connected to edge server \(x\), \(N_k\) is the total number of data used for training in the client \(k\). \(l_i^x\) is layer \(i\) from \(E_x^t\).

After aggregating the common layers from various edge-aggregated models, these global layers are combined to create several global models with heterogeneous architectures based on the requirements of different edge servers. The formula to combine the global layers is shown as:

\begin{equation}
    G_i^t = Gl_1^t \oplus Gl_2^t \oplus Gl_3^t \oplus \cdots \oplus Gl_i^t,
\end{equation}
where \(G_i^t\) is the global model for edge server \(i\) at round \(t\), and there are \(i\) layers in total. Algorithm \ref{alg:alg2} shows the aggregation process at the cloud servers. Table \ref{tab:my_label} shows notations used in our paper. 

\begin{algorithm}[h!]
    \setstretch{1.5}
    \caption{\textbf{The Aggregation at Cloud Server.} \(n\) is the number of layers contained by the edge server with the highest layer count. \(X_i\) is a set of edge servers whose aggregated models contain layer \(i\). \(E_x^t\) is an edge-aggregated model from edge server \(x\) at round \(t\). \(l_i^x\) is layer \(i\) from \(E_x^t\).  \(Gl_i^t\) is the global layer \(i\). \(K_x\) is the set of client devices connected to edge server \(x\). \(N_k\) is the total number of data used for training in the client \(k\). \(G_i^t\) is the global model for edge server \(i\). \(t\) is the round counter.}\label{alg:alg2}
    \begin{algorithmic}[1]
        \For{$i = 1,2,3,\dots,n$}
            \For {$x\in X_i$}
                \State $l_i^x = Extract(E_x,i)$
            \EndFor
            \State $ Gl_i^t = \sum_{x\in X_i} \dfrac{\sum_{k \in K_x} N_k}{\sum_{x\in X_i} \sum_{k \in K_x} N_k}l_i^x$
        \EndFor
        \For{$i = 1,2,3,\dots,n$}
           \State $G_i^t = Gl_1^t \oplus Gl_2^t \oplus Gl_3^t \oplus \cdots \oplus Gl_i^t$ 
        \EndFor
        \State \Return $G_1^t,G_2^t,G_3^t,\dots,G_n^t$
    \end{algorithmic}
\end{algorithm}
\vspace{-7mm}

\section{Implementation and Evaluation}
In our study, we evaluate the effectiveness of HAF-Edge using MNIST and FMNIST which are recognized as benchmarks in FL systems for image classification applications by comparing its performance with two baselines: vanilla FedAvg \cite{fl} and MaxCommon strategy \cite{flexifed}. The TensorFlow library of Python is used for implementation. In our experiment, we use heterogeneous models with different numbers of dense layers. Table.\ref {tab: fundation_model} represents the foundation model structure. The x-nn represents the model which has x number of dense layers except the output layer. As an example, 1nn and 5nn represent models that have 1 dense layer and 5 dense layers, respectively. To verify the effectiveness of our method, three different scenarios are applied to evaluate the performance of HAF-Edge. Table.\ref{tab:setting} shows the detailed settings of three experiment scenarios. 1NN and 3NN models are used in Scenario 1 and Scenario 2. In Scenario 3, 5 different model architectures from 1NN to 5NN are employed. To generate non-IID training sets for the client devices in three scenarios, MNIST and FMNIST datasets are sorted according to labels and partitioned into several subsets based on the required number of non-IID client devices in different scenarios, and each subset is regarded as one non-IID client. The client device with IID data contains all labels in three scenarios. 

\begin{table}[t]
\caption{Fundamental x-nn model architecture.}
\centering
\footnotesize\begin{tabular}{|l|c|c|c|}
\hline
\textbf{Layer} & \textbf{Output Shape} & \textbf{Activation} & \textbf{Parameters} \\
\hline
Input & (784,) & None & 0 \\
\hline
Dense & (200,) & ReLU & 157,000 \\
\hline
\multicolumn{4}{|c|}{\dots} \\
\hline
Dense & (10,) & Softmax & 2,010 \\
\hline
\end{tabular}

\label{tab: fundation_model}
\vspace{-3mm}
\end{table}

\begin{table}[b]
\vspace{-6mm}
     \caption{Basic settings of three scenarios}
    \centering
    \footnotesize\begin{tabular}{|c|c|c|c|c|}
    \hline
    Scenario & \makecell{Number of \\ Model Architectures} & \makecell{Client Number \\ per Edge} & \makecell{Client Data \\ Volume} & \makecell{IID Clients : Non-IID Clients \\ in Each Edge}\\
    \hline
    1 & 2 & 6 & 6000 & 1:5 \\
    \hline
    2 & 2 & 60 & 600 & 1:5 \\
    \hline
    3 & 5 & 11 & 1200 & 1:10 \\
    \hline
    \end{tabular}
    \label{tab:setting}
\end{table}

\subsection{Experiment Setup}
We use test accuracy and convergence speed as evaluation metrics. The test accuracy is the accuracy of the global model on the test set, and the convergence speed is the number of communication rounds needed for methods to reach the target test accuracy of the global model. FedAvg and MaxCommon strategy are used as baselines, where FedAvg is applied to different clusters of client devices grouped by their model architectures to obtain global models with various architectures, and MaxCommon strategy is applied to client devices without clustering, the common layers are extracted and aggregated to create heterogeneous global models.

\subsection{Evaluation Results}
\begin{figure}[h!]
\begin{subfigure}{0.5\textwidth}
\includegraphics[width=0.9\linewidth, height=36 mm]{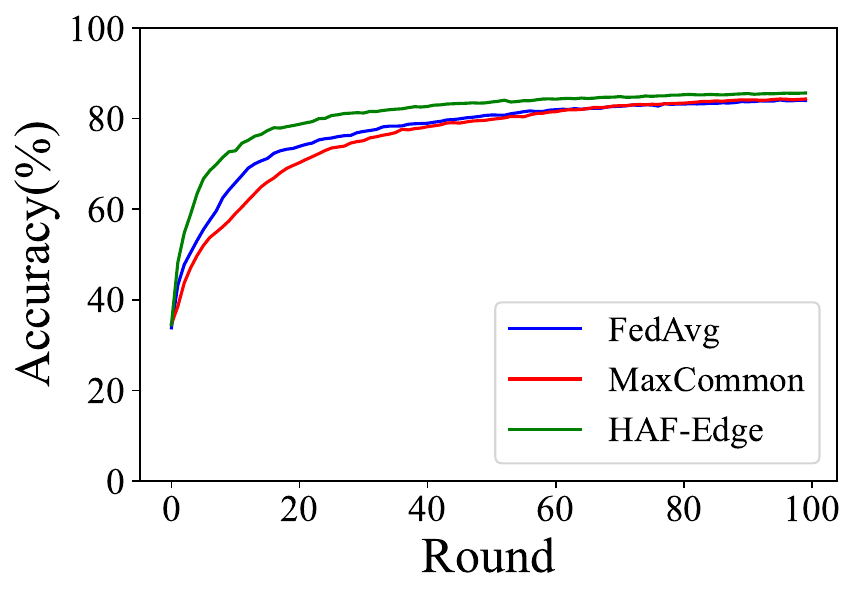} 
\caption{Global 1NN models}
\label{fig:MNIST_12clients_1NN}
\end{subfigure}
\begin{subfigure}{0.5\textwidth}
\includegraphics[width=0.9\linewidth, height=36 mm]{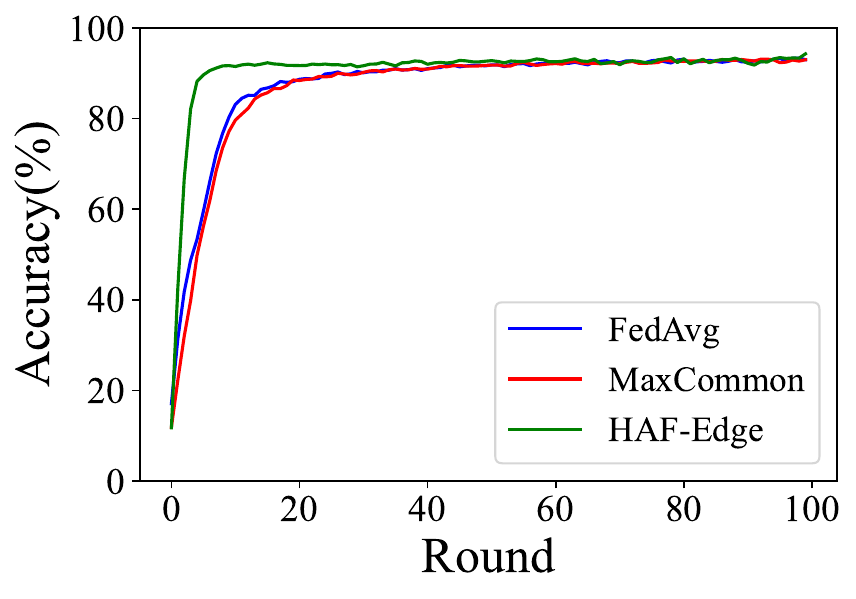}
\caption{Global 3NN models}
\label{fig:MNIST_12clients_3NN}
\end{subfigure}
\caption{Evaluating HAF-Edge, FedAvg, and MaxCommon strategy for Scenario 1 (2 edge servers, 6 clients for each edge server) using MNIST dataset.}
\label{fig:MNIST_12clients}
\end{figure}
\begin{figure}[h!]
\begin{subfigure}{0.5\textwidth}
\includegraphics[width=0.9\linewidth, height=36 mm]{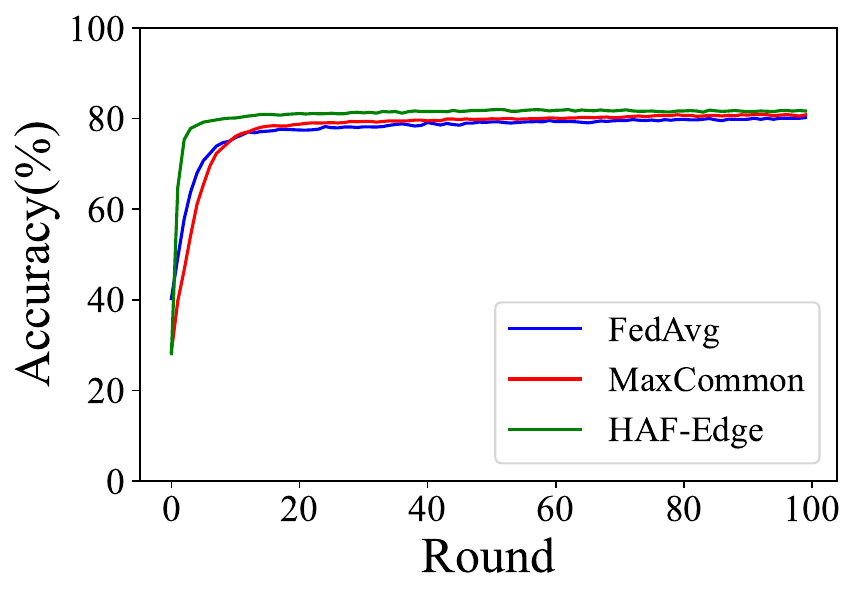} 
\caption{Global 1NN models}
\label{fig:FMNIST_12clients_1NN}
\end{subfigure}
\begin{subfigure}{0.5\textwidth}
\includegraphics[width=0.9\linewidth, height=36 mm]{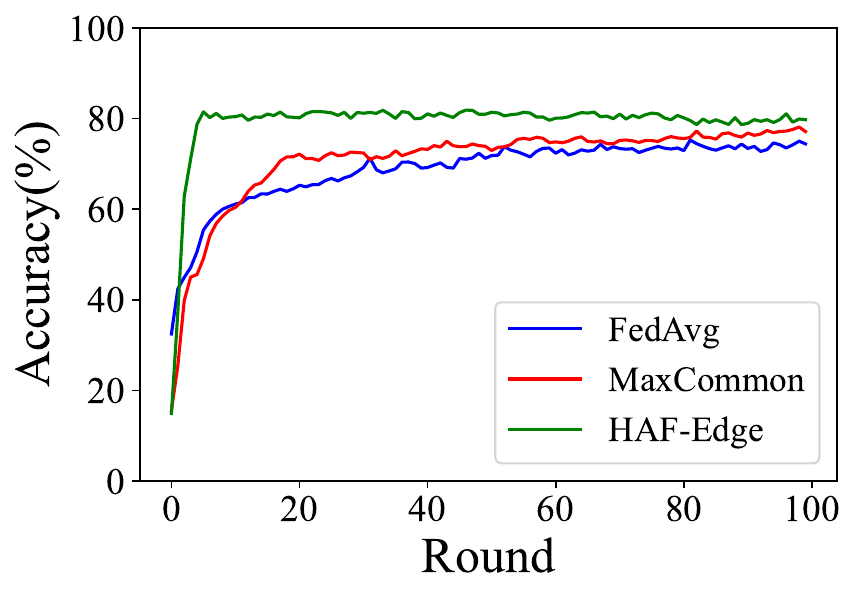}
\caption{Global 3NN models}
\label{fig:FMNIST_12clients_3NN}
\end{subfigure}
\caption{Evaluating HAF-Edge, FedAvg, and MaxCommon strategy for Scenario 1 (2 edge servers, 6 clients for each edge server) using FMNIST dataset.}
\label{fig:FMNIST_12clients}
\end{figure}
\begin{figure}[h!]
\begin{subfigure}{0.5\textwidth}
\includegraphics[width=0.9\linewidth, height=36 mm]{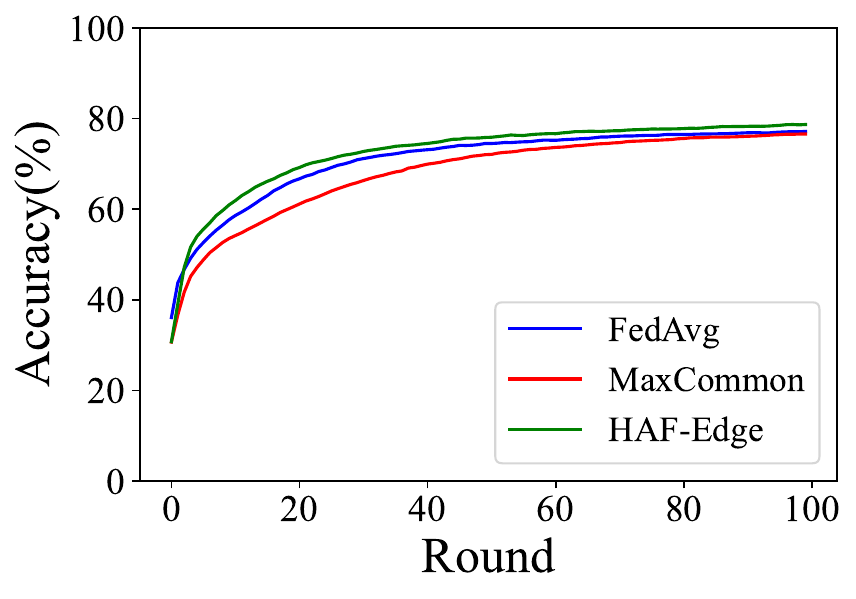} 
\caption{Global 1NN models}
\label{fig:MNIST_120clients_1NN}
\end{subfigure}
\begin{subfigure}{0.5\textwidth}
\includegraphics[width=0.9\linewidth, height=36 mm]{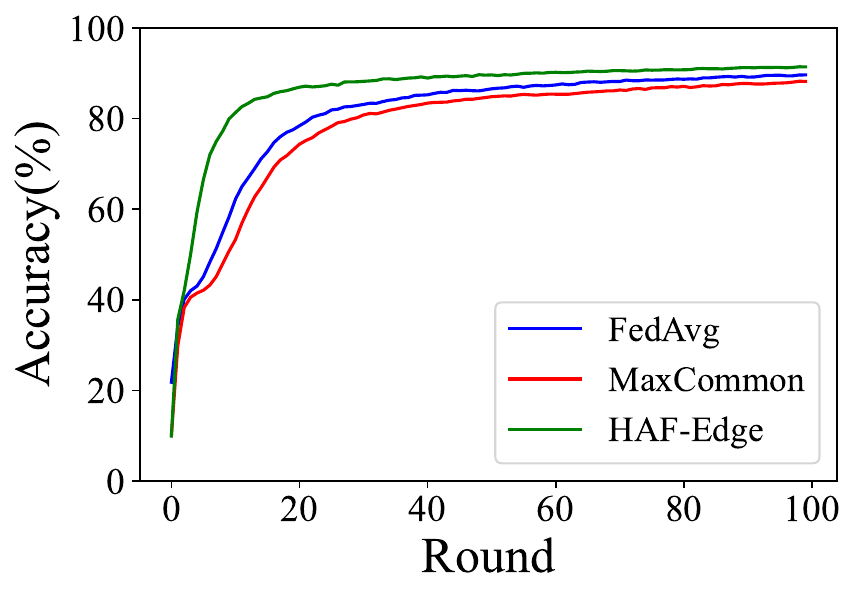}
\caption{Global 3NN models}
\label{fig:MNIST_120clients_3NN}
\end{subfigure}
\caption{Evaluating HAF-Edge, FedAvg, and MaxCommon strategy in Scenario 2 (2 edge servers, 60 clients for each edge server) with 1NN and 3NN models for the MNIST dataset.}
\label{fig:MNIST_120clients}
\end{figure}

\begin{figure}[h!]
\centering
\begin{subfigure}{0.5\textwidth}
    \centering
    \includegraphics[width=0.9\linewidth, height=36 mm]{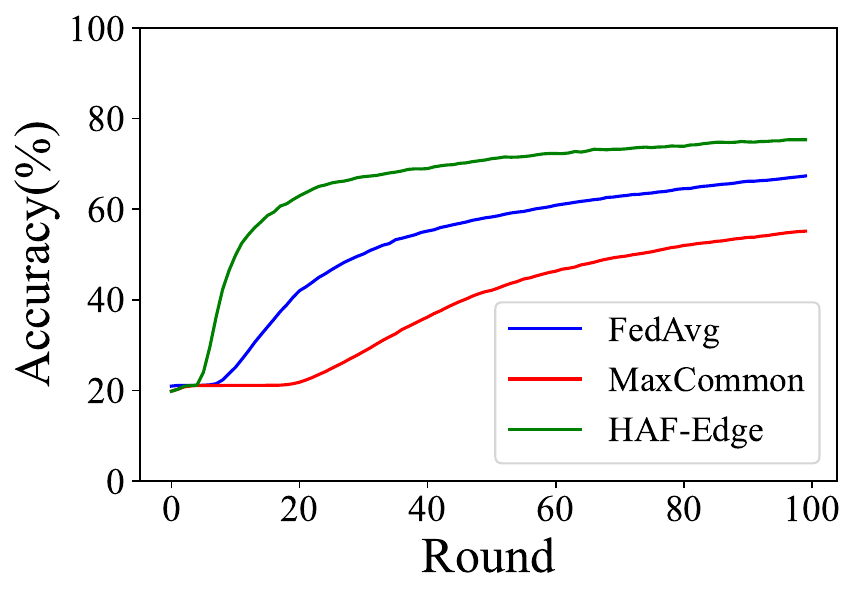} 
    \caption{Global 1NN models}
    \label{fig:MNIST_5edges_1NN}
\end{subfigure}%
\begin{subfigure}{0.5\textwidth}
    \centering
    \includegraphics[width=0.9\linewidth, height=36 mm]{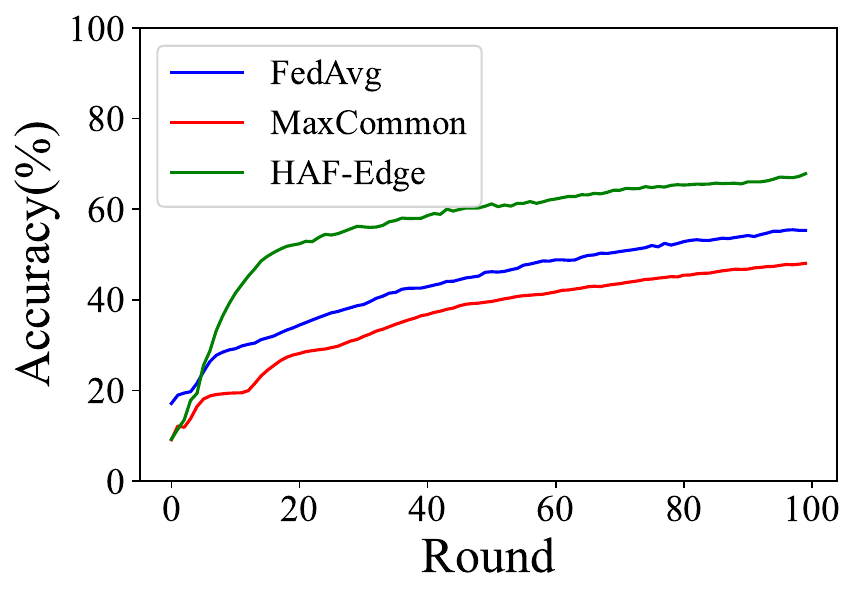}
    \caption{Global 2NN models}
    \label{fig:MNIST_5edges_2NN}
\end{subfigure}
\caption{Evaluating HAF-Edge, FedAvg, and MaxCommon strategy in Scenario 3 (5 edge servers, 11 clients for each edge server) with 1NN and 2NN models for the MNIST dataset.}
\label{fig:MNIST_5edges1}
\end{figure}

\begin{figure}[h!]
\begin{subfigure}{0.5\textwidth}
    \centering
    \includegraphics[width=0.9\linewidth, height=36 mm]{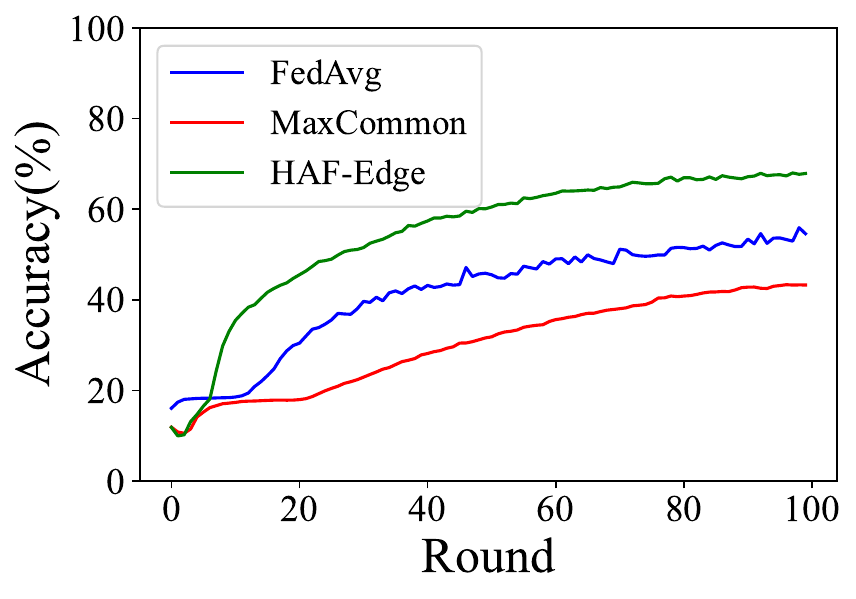}
    \caption{Global 3NN models}
    \label{fig:MNIST_5edges_3NN}
\end{subfigure}%
\begin{subfigure}{0.5\textwidth}
    \centering
    \includegraphics[width=0.9\linewidth, height=36 mm]{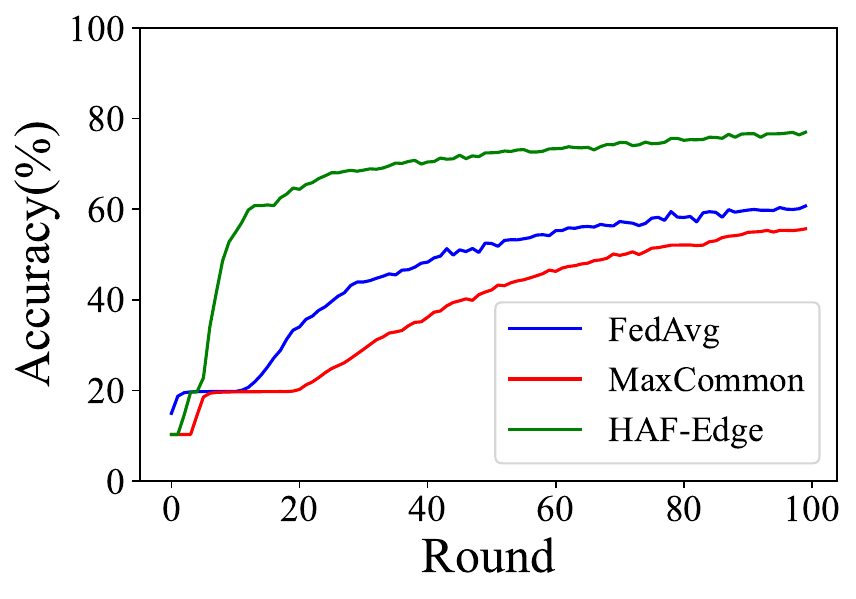}
    \caption{Global 4NN models}
    \label{fig:MNIST_5edges_4NN}
\end{subfigure}
\begin{subfigure}{1.0\textwidth}
    \centering
    \includegraphics[width=0.45\linewidth, height=36 mm]{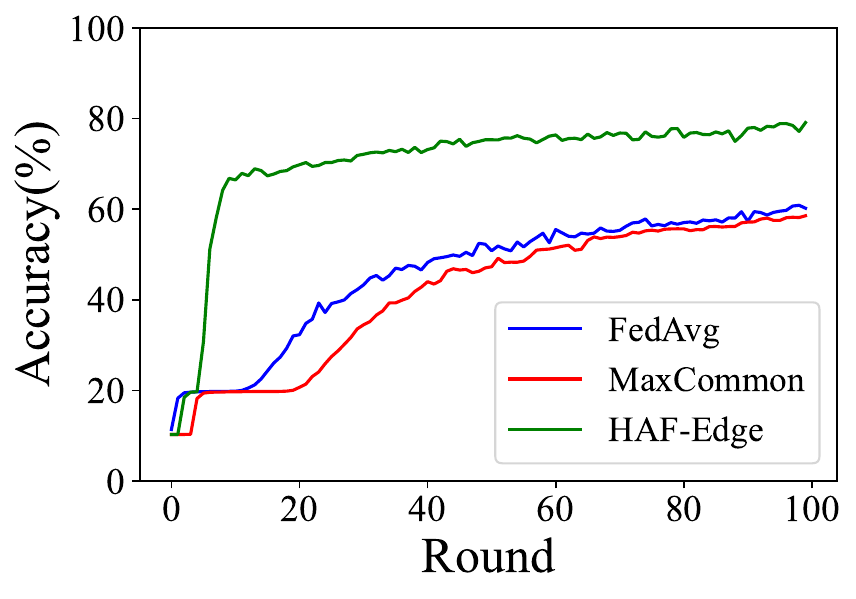}
    \caption{Global 5NN models}
    \label{fig:MNIST_5edges_5NN}
\end{subfigure}
\caption{Evaluating HAF-Edge, FedAvg, and MaxCommon strategy for Scenario 3 (5 edge servers, 11 clients for each edge server) with global 3NN, 4NN, and 5NN models.}
\label{fig:MNIST_5edges2}
\end{figure}
For Scenario 1, Figure \ref{fig:MNIST_12clients_1NN} and Figure \ref{fig:MNIST_12clients_3NN} present the test accuracy trends of global 1NN and 3NN models aggregated by HAF-Edge, FedAvg and MaxCommon strategy using MNIST dataset. The convergence speed of HAF-Edge is faster than that of FedAvg and MaxCommon strategy. The global 1NN model trained by HAF-Edge reaches \textbf{80\%} accuracy \textbf{25} rounds faster than FedAvg and \textbf{30} rounds faster than MaxCommon strategy. Moreover, HAF-Edge achieves \text{85\%} accuracy, whereas FedAvg and MaxCommon strategy only reaches 84\% with global 1NN models in 100 rounds. Figure \ref{fig:FMNIST_12clients_1NN} and Figure \ref{fig:FMNIST_12clients_3NN} show the test accuracy trends of global 1NN and 3NN models aggregated by HAF-Edge, FedAvg, and MaxCommon strategy using FMNIST dataset. HAF-Edge has a faster convergence speed and higher maximum accuracy on both global 1NN and 3NN models compared to FedAvg and MaxCommon strategy. The global 1NN model trained by HAF-Edge reaches \textbf{80\%} accuracy \textbf{73} rounds faster than FedAvg and \textbf{47} rounds faster than MaxCommon strategy. The maximum accuracy of the global 1NN model trained by HAF-Edge is around \textbf{1\%} higher than those trained by FedAvg and MaxCommon strategy in 100 rounds. The maximum accuracy of 3NN global models trained by FedAvg and MaxCommon strategy is \textbf{75\%} and \textbf{78\%} in 100 rounds, respectively. However, the maximum accuracy of the 3NN global model trained by HAF-Edge is \textbf{80\%}.

For Scenario 2, the test accuracy curves of global 1NN, and 3NN models aggregated by HAF-Edge, FedAvg, and MaxCommon strategy using the MNIST dataset are shown in Figure ~\ref{fig:MNIST_120clients}. Even though the number of client devices in each edge is increased compared to Scenario 1, the global 1NN model and 3NN model trained by HAF-Edge achieve better performance on both convergence speed and test accuracy. The maximum accuracy of the global 1NN model trained by HAF-Edge is \textbf{78\%}, which is \textbf{1\%} higher than that of FedAvg and \textbf{2\%} higher than that of  MaxCommon strategy. The global 3NN model trained by HAF-Edge achieves \textbf{91\%} accuracy, and global 3NN models trained by FedAvg and MaxCommon strategy reach \textbf{89\%} and \textbf{88\%} accuracy, respectively. 

For Scenario 3, the number of edge servers and model architecture types is increased compared to Scenario 1 and Scenario 2. Figures \ref{fig:MNIST_5edges1} and \ref{fig:MNIST_5edges2} present the test accuracy trends for 5 different global models (1NN, 2NN, 3NN, 4NN, 5NN). MNIST dataset is used in Scenario 3. The results of HAF-Edge are better in both test accuracy and convergence speed compared to the FedAvg and MaxCommon strategies. In all cases, the models trained by HAF-Edge can achieve at least \textbf{10\%} accuracy higher than models trained by FedAvg and MaxCommon. For the 5NN model, the maximum accuracy of HAF-Edge is about \textbf{80\%}, while the best performance of both FedAvg and MaxCommon reach to 60\% accuracy.

\section{Conclusion}

Non-IID data and model heterogeneity are two important challenges of FLaaS. In this paper, we present HAF-Edge, a three-level hierarchical federated learning where client devices with the same model settings are clustered and connected to an edge server that aims to solve these problems. In HAF-Edge, MaxCommon strategy is adopted at the cloud server to extract and aggregate common layers from models with different architectures. A distance-based weighting aggregation is proposed and applied at edge servers to reduce the impact of data non-IIDness, which is based on our observation that the Euclidean distance between local models trained with IID data and the global model is larger compared with the local models trained with non-IID data. We also evaluate the performance of HAF-Edge based on MNIST, FMNIST datasets under multiple scenarios. The results indicate that HAF-Edge outperforms FedAvg and MaxCommon Strategy under scenarios with non-IID data and heterogeneous models. 

\FloatBarrier  
\bibliographystyle{splncs04}
\bibliography{sca-fl.bib}

\begin{thebibliography}{10}
\providecommand{\url}[1]{\texttt{#1}}
\providecommand{\urlprefix}{URL }
\providecommand{\doi}[1]{https://doi.org/#1}

\bibitem{abad2020hierarchical}
Abad, M.S.H., Ozfatura, E., Gunduz, D., Ercetin, O.: Hierarchical federated learning across heterogeneous cellular networks. In: ICASSP 2020-2020 IEEE International Conference on Acoustics, Speech and Signal Processing (ICASSP). pp. 8866--8870. IEEE (2020)

\bibitem{healthcare}
Antunes, R.S., Andr{\'e}~da Costa, C., K{\"u}derle, A., Yari, I.A., Eskofier, B.: Federated learning for healthcare: Systematic review and architecture proposal. ACM Transactions on Intelligent Systems and Technology (TIST)  \textbf{13}(4),  1--23 (2022)

\bibitem{hierarchicalfl1}
Briggs, C., Fan, Z., Andras, P.: Federated learning with hierarchical clustering of local updates to improve training on non-iid data. In: 2020 international joint conference on neural networks (IJCNN). pp.~1--9. IEEE (2020)

\bibitem{chen2024optimizationfederatedlearningsclient}
Chen, S., Tavallaie, O., Hambali, M.H., Zandavi, S.M., Haddadi, H., Lane, N., Guo, S., Zomaya, A.Y.: Optimization of federated learning's client selection for non-iid data based on grey relational analysis (2024), \url{https://arxiv.org/abs/2310.08147}

\bibitem{chen2024rblarankbasedloraaggregationfinetuningheterogeneous}
Chen, S., Tavallaie, O., Nazemi, N., Zomaya, A.Y.: Rbla: Rank-based-lora-aggregation for fine-tuning heterogeneous models in flaas (2024), \url{https://arxiv.org/abs/2408.08699}

\bibitem{fair}
Deng, Y., Lyu, F., Ren, J., Chen, Y.C., Yang, P., Zhou, Y., Zhang, Y.: Fair: Quality-aware federated learning with precise user incentive and model aggregation. In: IEEE INFOCOM 2021-IEEE Conference on Computer Communications. pp. 1--10. IEEE (2021)

\bibitem{prediction}
Hard, A., Rao, K., Mathews, R., Ramaswamy, S., Beaufays, F., Augenstein, S., Eichner, H., Kiddon, C., Ramage, D.: Federated learning for mobile keyboard prediction. arXiv preprint arXiv:1811.03604  (2018)

\bibitem{hesamifard2018privacy}
Hesamifard, E., Takabi, H., Ghasemi, M., Wright, R.N.: Privacy-preserving machine learning as a service. Proceedings on Privacy Enhancing Technologies  (2018)

\bibitem{KnowledgeDistillation}
Hinton, G., Vinyals, O., Dean, J.: Distilling the knowledge in a neural network. arXiv preprint arXiv:1503.02531  (2015)

\bibitem{hsieh2020non}
Hsieh, K., Phanishayee, A., Mutlu, O., Gibbons, P.: The non-iid data quagmire of decentralized machine learning. In: International Conference on Machine Learning. pp. 4387--4398. PMLR (2020)

\bibitem{challenge1}
Kairouz, P., McMahan, H.B., Avent, B., Bellet, A., Bennis, M., Bhagoji, A.N., Bonawitz, K., Charles, Z., Cormode, G., Cummings, R., et~al.: Advances and open problems in federated learning. Foundations and trends{\textregistered} in machine learning  \textbf{14}(1--2),  1--210 (2021)

\bibitem{privcayconcern}
Kirienko, M., Sollini, M., Ninatti, G., Loiacono, D., Giacomello, E., Gozzi, N., Amigoni, F., Mainardi, L., Lanzi, P.L., Chiti, A.: Distributed learning: a reliable privacy-preserving strategy to change multicenter collaborations using ai. European Journal of Nuclear Medicine and Molecular Imaging  \textbf{48},  3791--3804 (2021)

\bibitem{flaas}
Kourtellis, N., Katevas, K., Perino, D.: Flaas: Federated learning as a service. In: Proceedings of the 1st workshop on distributed machine learning. pp. 7--13 (2020)

\bibitem{273723}
Lai, F., Zhu, X., Madhyastha, H.V., Chowdhury, M.: Oort: Efficient federated learning via guided participant selection. In: 15th {USENIX} Symposium on Operating Systems Design and Implementation ({OSDI} 21). pp. 19--35. {USENIX} Association (Jul 2021), \url{https://www.usenix.org/conference/osdi21/presentation/lai}

\bibitem{MNIST}
LeCun, Y., Cortes, C., Burges, C.J.: Mnist handwritten digit database  (2010), \url{http://yann.lecun.com/exdb/mnist/}

\bibitem{FedMD}
Li, D., Wang, J.: Fedmd: Heterogenous federated learning via model distillation. arXiv preprint arXiv:1910.03581  (2019)

\bibitem{challenge}
Li, T., Sahu, A.K., Talwalkar, A., Smith, V.: Federated learning: Challenges, methods, and future directions. IEEE Signal Processing Magazine  \textbf{37}(3),  50--60 (2020)

\bibitem{li2020federated}
Li, T., Sahu, A.K., Talwalkar, A., Smith, V.: Federated learning: Challenges, methods, and future directions. IEEE signal processing magazine  \textbf{37}(3),  50--60 (2020)

\bibitem{li2019convergence}
Li, X., Huang, K., Yang, W., Wang, S., Zhang, Z.: On the convergence of fedavg on non-iid data. arXiv preprint arXiv:1907.02189  (2019)

\bibitem{hierarchicalfl}
Liu, L., Zhang, J., Song, S., Letaief, K.B.: Client-edge-cloud hierarchical federated learning. In: ICC 2020 - 2020 IEEE International Conference on Communications (ICC). pp.~1--6 (2020)

\bibitem{tsc2}
Lu, X., Zheng, H., Liu, W., Jiang, Y., Wu, H.: Pop-fl: Towards efficient federated learning on edge using parallel over-parameterization. IEEE Transactions on Services Computing  \textbf{17}(2),  617--630 (2024)

\bibitem{fl}
McMahan, B., Moore, E., Ramage, D., Hampson, S., Arcas, B.A.y.: {Communication-Efficient Learning of Deep Networks from Decentralized Data}. In: Proceedings of the 20th ICAIS. vol.~54. PMLR (2017)

\bibitem{meng2022improving}
Meng, Q., Zhou, F., Ren, H., Feng, T., Liu, G., Lin, Y.: Improving federated learning face recognition via privacy-agnostic clusters. arXiv preprint arXiv:2201.12467  (2022)

\bibitem{nazemi_privacy}
Nazemi, N., Tavallaie, O., Chen, S., Mandalario, A.M., Thilakarathna, K., Holz, R., Haddadi, H., Zomaya, A.Y.: Access-fl: Agile communication and computation for efficient secure aggregation in stable federated learning networks. arXiv preprint arXiv:2409.01722  (2024)

\bibitem{tsc1}
Qu, Y., Yu, S., Gao, L., Sood, K., Xiang, Y.: Blockchained dual-asynchronous federated learning services for digital twin empowered edge-cloud continuum. IEEE Transactions on Services Computing  \textbf{17}(3),  836--849 (2024)

\bibitem{hierarchicalcommunication}
Rana, O., Spyridopoulos, T., Hudson, N., Baughman, M., Chard, K., Foster, I., Khan, A.: Hierarchical and decentralised federated learning. In: 2022 Cloud Continuum. pp.~1--9. IEEE (2022)

\bibitem{normdistance}
Smith, K.: Precalculus: A functional approach to graphing and problem solving. Jones \& Bartlett Publishers (2013)

\bibitem{flexifed}
Wang, K., He, Q., Chen, F., Chen, C., Huang, F., Jin, H., Yang, Y.: Flexifed: Personalized federated learning for edge clients with heterogeneous model architectures. In: Proceedings of the ACM Web Conference 2023. p. 2979–2990. WWW '23, Association for Computing Machinery, New York, NY, USA (2023)

\bibitem{276938}
Weng, Q., Xiao, W., Yu, Y., Wang, W., Wang, C., He, J., Li, Y., Zhang, L., Lin, W., Ding, Y.: {MLaaS} in the wild: Workload analysis and scheduling in {Large-Scale} heterogeneous {GPU} clusters. In: 19th USENIX Symposium on Networked Systems Design and Implementation (NSDI 22). pp. 945--960. USENIX Association, Renton, WA (Apr 2022), \url{https://www.usenix.org/conference/nsdi22/presentation/weng}

\bibitem{FMNIST}
Xiao, H., Rasul, K., Vollgraf, R.: Fashion-mnist: A novel image dataset for benchmarking machine learning algorithms. arXiv preprint arXiv:1708.07747  (2017), \url{https://arxiv.org/abs/1708.07747}

\bibitem{xu-etal-2023-federated}
Xu, Z., \textit{et al}.: Federated learning of gboard language models with differential privacy. In: Proceedings of the 61st Annual Meeting of the Association for Computational Linguistics (Volume 5: Industry Track). pp. 629--639. Association for Computational Linguistics (2023)

\bibitem{challenge2}
Ye, M., Fang, X., Du, B., Yuen, P.C., Tao, D.: Heterogeneous federated learning: State-of-the-art and research challenges. ACM Computing Surveys  \textbf{56}(3),  1--44 (2023)

\bibitem{non-iid}
Zhao, Y., Li, M., Lai, L., Suda, N., Civin, D., Chandra, V.: Federated learning with non-iid data. arXiv preprint arXiv:1806.00582  (2018)

\end{thebibliography}

\end{document}